\documentclass[a4paper,11pt]{article}
\usepackage{graphicx}%
\usepackage{amsmath,amssymb,amsfonts}%
\usepackage{amsthm}%
\usepackage{mathrsfs}%
\usepackage[top=2.5cm, bottom=2.5cm, left=1.8cm, right=1.8cm]{geometry}
\usepackage[title]{appendix}%
\usepackage{xcolor}%
\usepackage{textcomp}%
\usepackage{manyfoot}%
\usepackage{booktabs}%
\usepackage{algorithm}%
\usepackage{algorithmicx}%
\usepackage{algpseudocode}%
\usepackage{empheq}%
\usepackage{listings}%
\usepackage{hyperref}
\begin{document}
\title{\textsl{n}-dimensional Weyl conformal tensor in a four-index gravitational field theory}
\maketitle
\author{
\begin{center}
Frédéric Moulin \footnote{ frederic.moulin@ens-paris-saclay.fr} 
\end{center}} 
\noindent
\text{Université Paris-Saclay, Ecole Normale Supérieure Paris-Saclay} \\ 
\text{Département de Physique, 4 avenue des Sciences, 91190 Gif-sur-Yvette, France}
\begin{abstract}
The Weyl conformal tensor is the traceless component of the Riemann tensor and therefore, as is known, the information it contains does not appear explicitly in Einstein's equation. Following a rigorous mathematical treatment based on the variational principle, we will suggest that there exists a four-index gravitational field equation linearly containing the Weyl tensor closely related to a tidal gravitational field tensor whose components will be calculated.  The new degrees of freedom, introduced via the \textsl{n}-dimensional Weyl tensor, will therefore clearly appear as additional constraints on the metric and we will demonstrate, among other things, that the cosmological constant appears as a natural solution of the four-index theory in the form of an integration constant which therefore does not need to be introduced ad hoc into a Lagrangian. 
\end{abstract}
%
%
%
%
%
%
%
%
\section{Introduction}
The Riemann tensor $R_{ijkl}$ is the cornerstone of general relativity, but as everyone knows it does not appear explicitly in Einstein's equation. In a \textit{n=4} dimensional space, the 20 independent components of $R_{ijkl}$ distribute into the 10 components in the Ricci tensor $R_{jl}$ and 10 components in the Weyl tensor $C_{ijkl}$. This suggests that the 2-index Einstein equation may not be the most general equation and that an additional equation containing the Weyl tensor must naturally exist as we will demonstrate in this article. \\ 
The aim of this paper is to propose a four-index version of Einstein's equation linearly containing the Riemann tensor $R_{ijkl}$ and combinations of the Ricci tensor $R_{jl}$ and the Ricci scalar $R$,  as well as to solve it in well-known simple cases.  \\
The paper is organized as follows: \\
In section \ref{section2} of this article, using the principle of least action as well as a direct determination, we will demonstrate how to obtain this generalized 4-index equation. The right-hand side of this new equation will highlight the presence of an energy-momentum tensor of matter $T^{(M)}_{ijkl}$ as well as that of a tidal gravitational field tensor $T^{(F)}_{ijkl}$.\\
In part \ref{section2.4}, the generalized equation will be split into two distinct 4-index equations, the first containing the energy-momentum tensor of matter, and a second linked to the \textit{n}-dimensional Weyl conformal tensor $C_{ijkl}$ by the equation, $C_{ijkl}=-(n-3)\chi T^{(F)}_{ijkl}$. Due to the fact that the tensors $C_{ijkl}$ and $T^{(F)}_{ijkl}$ are both traceless, the information contained in the latter  equation is not contained in the 2-index Einstein equation.   
In a \textit{n}-dimensional space, the ${n(n+1)}/{2}$ components of the metric $g_{jl}$ are completely defined by the Einstein equation and we will see in part \ref{section2.5}, that the $n(n+1)(n+2)(n-3)/12$ new degrees of freedom, introduced via the  \textit{n}-dimensional Weyl tensor, will appear as additional constraints on the metric. 
In part \ref{section2.7}, we will determine the structure of the tidal gravitational field tensor $T^{(F)}_{ijkl}$. \\
In section \ref{section3}, the main result of this paper will be given by the resolution of this new four-index equation applied to the case of the Schwarzschild metric in a \textit{n=4} dimensional space.
We will show, among other things, that the solution for the gravitational field potential naturally contain an expected first term produced by the presence of matter, $(r_s/r)$, but also, at the same time, a second term, $(\lambda r^2/3)$, where $\lambda$ appears as a simple integration constant that hasn't been introduced ad hoc into the 4-index equation or into the Lagrangian.
The additional degrees of freedom, introduced through the 4-index theory, clearly show us that the integration constant $\lambda$, appears as an additional constraint probably closely linked to a still unknown property of the gravitational field.
The comparison with the famous cosmological constant $\Lambda$ will be discussed in \ref{section3.2}, and the problem of its usual physical interpretation in terms of dark energy does not appear to be the one highlighted in this article. \\
In the final section \ref{section4}, a brief study shows that the 4-index equation is also compatible with the Friedmann-Lemaître-Robertson-Walker (FLRW) metric.
\section{Generalized four-index gravitational field equation}  \label{section2}
In this section, the principle of least action will be used to obtain the generalized 4-index version of Einstein's equation.
    \subsection{Lagrangian formulation in \textit{n}-dimensions} \label{section2.1}
The formulation of the Lagrangian for general relativity $L^{(G)}$, is simply given by the Ricci scalar $R$ allowing us to write the well-known Einstein-Hilbert action  $S^{(G)}$ as \cite{citLandau}:
\begin{equation}
S^{(G)} = -\frac{1}{2 \chi c }\int L^{(G)} \sqrt{-g} \,d\Omega=-\frac{1}{2 \chi c }\int R \sqrt{-g} \,d\Omega, \label{EinsteinHilbert}
\end{equation}
with the Einstein constant $\chi=8 \pi G/c^{4}$. \\
As stated in many works \cite{citLandau,citMisner,citBlau,citOyv}, it is well known that to obtain the famous 2-index Einstein field equation, it is necessary to introduce the Ricci tensor $R_{jl}$ via the relation $R = g^{jl}R_{jl}$: 
\begin{equation}
S^{(G)} = -\frac{1}{2 \chi c }\int g^{jl}R_{jl} \sqrt{-g} \,d\Omega.
\end{equation} 
If we now want to find the 4-index version of Einstein's equation, while remaining within the framework of general relativity, we need to introduce a new tensor $L_{ijkl}$ such that:
\begin{equation}
 R = g^{jl}R_{jl} = g^{jl} g^{ik} L_{ijkl}. \label{lagrangienLijkl}
\end{equation}
and then:
\begin{equation}
S^{(G)} = -\frac{1}{2 \chi c }\int g^{jl} g^{ik} L_{ijkl}\sqrt{-g} \,d\Omega. \label{SGLijkl}
\end{equation} 
To have the most general formulation possible, this tensor $L_{ijkl}$ must contain linearly not only the Riemann tensor $R_{ijkl}$, but also two other expressions constructed from combinations containing $R_{jl}$ and $R$ and exhibiting the same symmetries as $R_{ijkl}$. It is easy to show that the following three expressions have exactly the same desired symmetries: 
\begin{align}
&(R_{ijkl}),    \label{Riemann} \\
&(g_{ik}R_{jl}-g_{jk}R_{il}+g_{jl}R_{ik}-g_{il}R_{jk}),   \label{gijklRik} \\
&(g_{ik}g_{jl}-g_{il}g_{jk})R.   \label{gijklR} 
\end{align}
The general tensor $L_{ijkl}$ can therefore be written as a linear combination of these three tensors:
\begin{equation}
 L_{ijkl} =  a R_{ijkl}  + b (g_{ik}R_{jl}-g_{jk}R_{il}+g_{jl}R_{ik}-g_{il}R_{jk})  + d (g_{ik}g_{jl}-g_{il}g_{jk})R, \label{Lijkl} 
\end{equation}
where $a$, $b$, $d$ are three arbitrary parameters that can be determined by contraction and identification with relation (\ref{lagrangienLijkl}) and with $g^{jl}g_{jl}=\delta_{j}^{j}=n$ in a \textit{n}-dimensional space:
\begin{equation}
b=\frac{(1-a)}{(n-2)} \quad , \quad   d=-\frac{(1-a)}{(n-1)(n-2)} \label{abd} 
\end{equation}
allowing us to rewrite $L_{ijkl}$ in a form that involves only one parameter $a$: 
\begin{equation}
 L_{ijkl} =  a R_{ijkl}  + \frac{(1-a)}{(n-2)}\, (g_{ik}R_{jl}-g_{jk}R_{il}+g_{jl}R_{ik}-g_{il}R_{jk}) 
 - \frac{(1-a)}{(n-1)(n-2)}\, (g_{ik}g_{jl}-g_{il}g_{jk})R, \label{Lijkl2} 
\end{equation}
which corroborates well with:
\begin{align}
  g^{ik} L_{ijkl} &= R_{jl},    \label{contractLijkl1} \\
  g^{jl} g^{ik} L_{ijkl} &= R,   \label{contractLijkl}
\end{align}	
whatever the values of $a$ and $n$, as well as the expected symmetries: 
\begin{equation}
L_{ijkl} =-L_{jikl}=-L_{ijlk}=L_{klij}.  \label{symLijkl}
\end{equation} 
As shown in (\ref{nablaGijkl}), as well as in a previous paper \cite{citMoulin}, the parameter $a$ must take a particular value, $a=-1/(n-3)$, which is the necessary condition ensuring conservation of the total energy-momentum tensor. 
The relation (\ref{Lijkl2}) for $L_{ijkl}$, written with $a=-1/(n-3)$, will be used subsequently:
\begin{equation}
 L_{ijkl} =   \frac{1}{(n-3)}\, \big[ - R_{ijkl}   + (g_{ik}R_{jl}-g_{jk}R_{il}+g_{jl}R_{ik}-g_{il}R_{jk})  
 - \frac{1}{(n-1)}\, (g_{ik}g_{jl}-g_{il}g_{jk})R \, \big]. \label{Lijkl} 
\end{equation}
   \subsection{4-index version of Einstein's equation} \label{section2.2}
The mathematical formulas required here for the principle of least action will be exactly  the same as those usually used to obtain the well-known 2-index Einstein field equation.
The variational principle has the following form:
\begin{equation}
\delta S = 0, \label{deltaS}
\end{equation}
where the total action $S$ is the sum of the purely gravitational Einstein-Hilbert action $S^{(G)}$ defined in (\ref{EinsteinHilbert}), and the total matter-field action $S^{(MF)}$ \cite{citLandau}:    
\begin{equation}
 S^{(MF)} = \frac{1}{c}\int L^{(MF)}\sqrt{-g} \,d\Omega, \label{actiontotale}
\end{equation}
written with the Lagrangian $L^{(MF)}$ which describes all matter, energy and fields present in the spacetime. \\ 
We now need to determine the variation of the total action:
\begin{equation}
\delta S = \delta S^{(G)} \,+\, \delta S^{(MF)}   
  = -\frac{1}{2 \chi c}\, \delta\int L^{(G)} \sqrt{-g} \,d\Omega \,+\, \frac{1}{c} \delta\int L^{(MF)}\sqrt{-g} \,d\Omega. \label{varactiontotale}
\end{equation} 
First, let us calculate the first term, $\delta S^{(G)}$, using the relation (\ref{SGLijkl}): 
\begin{align}
 \delta S^{(G)}
             &= -\frac{1}{2 \chi c }\int \delta \big[ \, g^{jl}g^{ik}L_{ijkl} \sqrt{-g}  \,\big]  \, d\Omega \nonumber \\
						 &=-\frac{1}{2 \chi c}\int \big[\, \delta g^{jl} g^{ik}L_{ijkl}  \sqrt{-g}  
						 + g^{jl} \delta (g^{ik}L_{ijkl}) \sqrt{-g}   
						   +   g^{jl}g^{ik}L_{ijkl} \, \delta \sqrt{-g} \,\big]    d\Omega. \label{deltaLLijkl}
\end{align}
The symbolic notation $\delta$ describes the variation with respect to the components of the metric tensor as well as its derivatives, and  we will use the two well-known formulas written below and demonstrated in many references \cite{citLandau,citMisner,citBlau,citOyv}:
\begin{align}
 & \int g^{jl}\delta R_{jl} \, \sqrt{-g} \, \, d\Omega =0,         \label{deltagjlRjl} \\
& \delta \sqrt{-g}=-\frac{1}{2} \,g_{jl}\, \delta g^{jl}\,\sqrt{-g}.   \label{deltaLgjl}  
\end{align}
Using the relations (\ref{contractLijkl1}) and (\ref{deltagjlRjl}), we can see that the second term of (\ref{deltaLLijkl}) cancels out:
\begin{equation}
\int  g^{jl} \delta (g^{ik}L_{ijkl}) \sqrt{-g} \, d\Omega = \int g^{jl}\delta R_{jl} \, \sqrt{-g}  \, d\Omega =0.
\end{equation} 
Using the relations (\ref{contractLijkl}) and (\ref{deltaLgjl}), we notice that the third term of (\ref{deltaLLijkl}) transforms into:
\begin{equation}
\int  g^{jl}g^{ik}L_{ijkl} \, \delta \sqrt{-g} \, d\Omega = 
  -\int \frac{1}{2 } \, g_{jl} \, \delta g^{jl} \, R \sqrt{-g}  \, d\Omega,   \label{thirdterm} 
\end{equation}
and in order to keep the same symmetries (\ref{symLijkl}) as for $L_{ijkl}$, we can use the following mathematical trick:
\begin{equation}
g_{jl} =  \frac{1}{(n-1)} \, g^{ik} (g_{ik}g_{jl}-g_{il}g_{jk}), \label{tricks}
\end{equation}
allowing  us to rewrite (\ref{thirdterm}) in an equivalent mathematical form but this time with the expected symmetries:
\begin{equation}
-\int \frac{1}{2(n-1) }\, g^{ik} (g_{ik}g_{jl}-g_{il}g_{jk}) \, \delta g^{jl}  R \sqrt{-g}  \, d\Omega.   
\end{equation}
Finally, $\delta S^{(G)}$ becomes:
\begin{equation}
 \delta S^{(G)} =  -\frac{1}{2 \chi c }\int  g^{ik} \big[ \, L_{ijkl}  - \frac{1}{2(n-1) } \,(g_{ik}g_{jl}-g_{il}g_{jk})R \,\big]\, \delta g^{jl} \sqrt{-g} \, d\Omega.  
\end{equation}
For the part concerning the variation $\delta \, L^{(MF)}$, we will first write the usual definition of the total energy-momentum tensor $T_{jl}$ \cite{citLandau}:
\begin{equation}
\frac{1}{2} \, \sqrt{-g} \, T_{jl}=\frac{\delta \, ( \sqrt{-g} \, L^{(MF)} ) }{\delta g^{jl} },  \label{Tjldelta}
\end{equation}
and, if we introduce a 4-index generalized energy-momentum tensor $T_{ijkl}$, whose contraction gives:
\begin{equation}
 g^{ik} T_{ijkl}=T_{jl},  \label{gikTijkl}
\end{equation}
we can calculate the variation $\delta S^{(MF)}$:
\begin{align}
 \delta S^{(MF)} &= \frac{1}{ c }\int \delta \big[\,  L^{(MF)} \sqrt{-g}\,\big] \, d\Omega   \nonumber \\
						& =  \frac{1}{2c}\, \int  T_{jl}\,   \delta g^{jl}  \sqrt{-g} \, d\Omega  \nonumber \\
						& =  \frac{1}{2c}\, \int     g^{ik} T_{ijkl}\,    \delta g^{jl}  \sqrt{-g} \, d\Omega. 
\end{align}
The variation of the total action $\delta S$ finally gives:
\begin{align}
 \delta S  =&\, \delta S^{(G)} \, + \, \delta S^{(MF)}   \nonumber \\
           =& -\frac{1}{2 \chi c}\,\int   g^{ik} \big[ \,L_{ijkl} - \frac{1}{2(n-1)}(g_{ik}g_{jl}-g_{il}g_{jk})R \, \big] \, \delta g^{jl} \sqrt{-g} \, d\Omega     \nonumber \\
	    & +\frac{1}{2  c}\,\int     g^{ik} T_{ijkl}  \,\delta g^{jl} \sqrt{-g} \, d\Omega.
\end{align} 
Now, considering that the total action, $\delta S=\delta S^{(G)}+\delta S^{(MF)}=0$,  should hold for any variation of $\delta g^{jl}$, we can write:
\begin{equation}
g^{ik} \big[ \,L_{ijkl} - \frac{1}{2(n-1)}(g_{ik}g_{jl}-g_{il}g_{jk})R  \, \big]=g^{ik}  \big[ \, \chi T_{ijkl} \, \big].   \label{gikLijklbis}
\end{equation}
The principle of least action used here is obviously the same as that usually used in general relativity, and the result of the above contraction with, $g^{ik}L_{ijkl}=R_{jl}$ (\ref{contractLijkl1}), $g^{ik}(g_{ik}g_{jl}-g_{il}g_{jk})=(n-1)g_{jl}$ (\ref{tricks}) and with the definition, $g^{ik} T_{ijkl}=T_{jl}$ (\ref{gikTijkl}), gives back, as expected, the famous 2-index Einstein equation  whatever the value of \textit{n}:
\begin{equation}
R_{jl}-\frac{1}{2}\,g_{jl}R= \chi T_{jl}, \label{eqeinstein}
\end{equation}
emphasizing that we did not use the above equation in any intermediate calculations. \\
In general, for arbitrary $T_{ijkl}$, we can not simply exclude the contractional factor $g^{ik}$ in (\ref{gikLijklbis}). However, it is always possible to choose a energy-momentum tensor $T_{ijkl}$ which has the same number of components and the same symmetries as $L_{ijkl}$ (or $R_{ijkl}$), and which acts as a source term,  in the same way as the tensor $T_{jl}$, allowing us to write by identification the 4-index version of Einstein's equation in a \textit{n}-dimensional space:
\begin{equation}
 L_{ijkl} - \frac{1}{2(n-1)}(g_{ik}g_{jl}-g_{il}g_{jk})R =  \chi T_{ijkl}. \label{LijklT}
\end{equation}
This new equation contains the Riemann tensor and therefore the traceless Weyl tensor, which, as everyone knows, does not appear explicitly in Einstein's equation. 
Here we find the ${n^2(n^2-1)}/{12}$ degrees of freedom associated with $R_{ijkl}$, compared with the ${n(n+1)}/{2}$ associated with $R_{jl}$ (see part \ref{section2.5}), so, despite being a 4-index version, the general equation (\ref{LijklT}) contains more information than (\ref{eqeinstein}) and we will see later the interesting consequences that this formulation can bring to the theory of general relativity. The $n(n+1)(n+2)(n-3)/12$ $(=n^2(n^2-1)/12-n(n+1)/2)$  degrees of freedom of differences are exactly provided by the Weyl tensor (part \ref{section2.5}) and therefore lost during the contraction in (\ref{gikLijklbis}). \\
By replacing $L_{ijkl}$ with its expression (\ref{Lijkl}), in (\ref{LijklT}), we obtain the generalized 4-index equation:  
\begin{equation}
\frac{1}{(n-3)} [-R_{ijkl}  + (g_{ik}R_{jl}-g_{jk}R_{il}+g_{jl}R_{ik}-g_{il}R_{jk})  
 - \frac{1}{2}  (g_{ik}g_{jl}-g_{il}g_{jk})R]=\chi T_{ijkl} \label{eqdual4index1}
\end{equation}
where we recognize here the well known expression of the Riemann double dual tensor  given by \cite{citMisner}:
\begin{align}
 ^{*}R_{ijkl}^{*}  &= \frac{1}{4}\, e_{ijpq}\, R^{pqrs}e_{klrs}  \nonumber \\
&= -R_{ijkl}  + (g_{ik}R_{jl}-g_{jk}R_{il}+g_{jl}R_{ik}-g_{il}R_{jk} )  
- \frac{1}{2} \, (g_{ik}g_{jl}-g_{il}g_{jk})R, \label{doubledualRiemann}
\end{align}  
where $e_{ijkl}$ is the Levi-Civita tensor. \\
The general 4-index Einstein's equation can thus be written in a more compact form:
\begin{equation}
 ^{*}R_{ijkl}^{*} = \chi (n-3) \, T_{ijkl}.  \label{eqdual4index} 
\end{equation}  
We can easily verify that the tensorial contraction of this equation, with $g^{ik} {} ^{*}R_{ijkl}^{*}= (n-3) (R_{jl}-1/2\,g_{jl}R)$, of course gives Einstein's equation (\ref{eqeinstein}) whatever the value of \textit{n},  meaning, incidentally, that the 2-index Einstein equation therefore has the same form in any space-time of dimension \textit{n}, which is also an important result.
		\subsection{Energy-momentum conservation}  \label{section2.3}
The contracted Bianchi identities, $\nabla_{i} R^{i}{}_{jkl}=\nabla_{k} R_{jl}-\nabla_{l} R_{jk}$ and $\nabla_{j} R^{j}{}_{l}=1/2 \,\nabla_{l} R$, allow us to calculate the covariant derivative:   \\
\begin{equation}
\nabla_{i}{}^{*}R^{*}{}^{i}{}_{jkl}=0,  \label{nablaDRDijkl}
\end{equation}
ensuring the conservation of the total energy-momentum tensor:
\begin{equation}
\nabla_{i}\, T^{i}{}_{jkl}=0,  \label{nablaTijkl}
\end{equation}
which justifies the choice of the tensor $L_{ijkl}$ in (\ref{Lijkl}), write with the parameter $a=-1/(n-3)$. 
The equation (\ref{eqdual4index}) is indeed the 4-index version of Einstein's equation and fits perfectly into the framework of general relativity.
   \subsection{Two-part decomposition of the 4-index equation}   \label{section2.4}
It is well known that Einstein himself and a major consensus of famous physicists have emphasized that the gravitational field must also have an energy-momentum tensor as do all other physical fields. In this section, we will show that the 4-index gravitational field equation (\ref{eqdual4index1}) can be split into two parts, depending on whether we're considering matter $^{(M)}$ or gravitational field ${}^{(F)}$. \\
It is possible to decompose the Lagrangian of the matter-field into two different parts, $L^{(MF)}=L^{(M)}+L^{(F)}$, thus allowing us to divide the general tensor $T_{ijkl}$ also into two parts, one represented by the energy-momentum tensor of matter $T^{(M)}_{ijkl}$ present in space-time, and the other by a tensor $T^{(F)}_{ijkl}$ associated with the gravitational field:
\begin{equation}
T_{ijkl}=T^{(M)}_{ijkl}+T^{(F)}_{ijkl}. \label{TijklTMTF}  
\end{equation}
We will demonstrate later that $T^{(F)}_{ijkl}$ is rather associated with the tidal gravitational field tensor and it is well known that a free-falling observer may assume that he is not in the presence of a gravitational field, but a tidal gravitational field cannot be cancelled out by a choice of reference system which also represents the whole interest of this tensor decomposition. We will also show that $T^{(F)}_{ijkl}$ is a traceless tensor. \\
The Riemann tensor and the \textsl{n}-dimensional Weyl tensor $C_{ijkl}$, are connected by the well-known formula \cite{citBlau, citOyv}:
\begin{equation}
C_{ijkl} = R_{ijkl}-\frac{1}{(n-2)}(g_{ik}R_{jl}-g_{jk}R_{il}+g_{jl}R_{ik} -g_{il}R_{jk})  
     + \frac{1}{(n-1)(n-2)}   (g_{ik}g_{jl}-g_{il}g_{jk})R,     \label{Weytensor} 
\end{equation} 
from which we can replace the Riemann tensor $R_{ijkl}$ into the general equation (\ref{eqdual4index1}), with (\ref{TijklTMTF}) to obtain: 
\begin{equation}
B_{ijkl} - \frac{1}{(n-3)} \, C_{ijkl}   = \chi (T^{(M)}_{ijkl}+T^{(F)}_{ijkl}),   \label{CijklBijkl}
\end{equation}
with a new tensor $B_{ijkl}$ given by the formula:
\begin{equation}
B_{ijkl}=  \frac{1}{(n-2)} \,(g_{ik}R_{jl}-g_{jk}R_{il}+g_{jl}R_{ik}-g_{il}R_{jk})  
  -\frac{n}{2(n-1)(n-2)} \, (g_{ik}g_{jl}-g_{il}g_{jk}) R.   \label{Bijkl}  
\end{equation}
The tensorial contraction of the Einstein equation (\ref{eqeinstein}), gives the Ricci scalar $R$ in the form: 
\begin{equation}
R=-\frac{2}{(n-2)}\,\chi \,T, \label{R} 
\end{equation}
and so the Ricci tensor $R_{jl}$: 
\begin{equation}
R_{jl}=\chi \, (T_{jl}-\frac{1}{(n-2)}\,g_{jl}T), \label{Rjl}
\end{equation}
which we now substitute into (\ref{Bijkl}), obtaining:
\begin{equation}
B_{ijkl}= \chi [\frac{1}{(n-2)}  (g_{ik}T_{jl}-g_{jk}T_{il}+g_{jl}T_{ik} -g_{il}T_{jk}) 
 - \frac{1}{(n-1)(n-2)}  (g_{ik}g_{jl}-g_{il}g_{jk})T].    \label{BijklTjl}
\end{equation} 
The above relationship indicate that the tensor $B_{ijkl}$ is directly linked to the energy-momentum tensor of the matter content present in the standard theory of general relativity by means of the tensor $T_{jl}$ and $T$.
The right-hand side of (\ref{BijklTjl}) represents a 4-index energy-momentum tensor for matter that can only be identified with  $T^{(M)}_{ijkl}$ in (\ref{CijklBijkl}), which can be written as follows:  
\begin{equation}
T^{(M)}_{ijkl} =   \frac{1}{(n-2)} \, (g_{ik}T_{jl}-g_{jk}T_{il}+g_{jl}T_{ik} -g_{il}T_{jk})   
- \frac{1}{(n-1)(n-2)}  \, (g_{ik}g_{jl}-g_{il}g_{jk})\,T.    \label{TMijkl}  
\end{equation}
The remaining traceless part in equation (\ref{CijklBijkl}), is therefore given by the Weyl tensor which is consequently simply related to $T^{(F)}_{ijkl}$ (it is well known that the Weyl tensor encodes information about the gravitational field in free space devoid of matter).
The generalized equation (\ref{eqdual4index1}) can now be written in the equivalent form of two 4-index equations:
\begin{align}
  B_{ijkl} & =  \chi  T^{(M)}_{ijkl}  \label{BijklTMijkl} \\
  C_{ijkl} &  = -\chi (n-3)  T^{(F)}_{ijkl}     \label{CijklTFijkl}
\end{align}
These two equations are coupled via the metric $g_{jl}$ and its derivatives and in a \textit{n}-dimensional space, we can use either the equation (\ref{eqdual4index1}) or the two equations (\ref{BijklTMijkl}) and (\ref{CijklTFijkl}). 
The tensor $T^{(F)}_{ijkl}$ is not directly determined by the matter content and therefore it does not vanish into the vacuum either.
Its structure will be determined later in the section (\ref{section2.7}).  \\
We note that $B_{ijkl}$ and $T^{(M)}_{ijkl}$ ultimately involve only the tensors $R_{jl}$, $R$, $T_{jl}$ and $T$ and it is easy to verify that the contraction of (\ref{BijklTMijkl}) gives the Einstein equation (\ref{eqeinstein}). 
We can also show, by verifying it on many known metrics, that for the case of a \textit{n=4} dimensional space-time, the equation (\ref{BijklTMijkl}) yields exactly the same physical solutions as those classically obtained with Einstein's equation. 
This result is not entirely surprising given the identical number of independent components for the $B_{ijkl}$ and $R_{jl}$ tensors, as we shall see in section \ref{section2.5}.  This is of course an important result, enabling us to check our calculations at this stage, while giving physical meaning to the two-part decomposition.  
   \subsection{Direct determination of the four-index gravitational field equation}
In this part, we will show that it is also possible to find the general equation (\ref{eqdual4index}), as well as the two equations (\ref{BijklTMijkl}) and (\ref{CijklTFijkl}), quite easily, using a direct method without going through a principle of least action. Let's start by writing a generalized four-index gravitational field equation in the form: 
\begin{equation}
 G_{ijkl} = \chi \, T_{ijkl},  \label{GijklTijkl} 
\end{equation}  
such that the contraction gives back the Einstein equation (\ref{eqeinstein}):
\begin{align}
 g^{ik}G_{ijkl}& = G_{jl} = R_{jl}-\frac{1}{2}\,g_{jl}R, \label{contractGijkl} \\
g^{ik}T_{ijkl}& = T_{jl}.  \label{contractTijkl}
\end{align}
Just like the Riemann tensor $R_{ijkl}$, the generalized fourth-order Einstein tensor $G_{ijkl}$ must also contain a trace-free part corresponding to the Weyl tensor $C_{ijkl}$ (\ref{Weytensor}), as well  as linear combinations involving $R_{jl}$ and $R$ already defined in (\ref{gijklRik}) and (\ref{gijklR}). By analogy, we can also write $T_{ijkl}$ with a trace-free part noted $T^{(C)}_{ijkl}$, as well as linear combinations involving $T_{jl}$ and $T$:
\begin{equation}
 G_{ijkl} =  a C_{ijkl}  + a_1 (g_{ik}R_{jl}-g_{jk}R_{il}+g_{jl}R_{ik}-g_{il}R_{jk})  + a_2 (g_{ik}g_{jl}-g_{il}g_{jk})R, \label{Gijkl}
\end{equation}
\begin{equation}
T_{ijkl} =  e T^{(C)}_{ijkl}  + e_1 (g_{ik}T_{jl}-g_{jk}T_{il}+g_{jl}T_{ik}-g_{il}T_{jk})  + e_2 (g_{ik}g_{jl}-g_{il}g_{jk})T, \label{TCijkl} 
\end{equation}
where $a$, $a_1$, $a_2$, $e$, $e_1$, $e_2$ are arbitrary parameters. \\
The contractions (\ref{contractGijkl}) and (\ref{contractTijkl}) allows us to determine $a_1$, $a_2$ and $e_1$, $e_2$  such that:
\begin{align}
&a_1=\frac{1}{(n-2)} \quad , \quad   a_2=-\frac{n}{2(n-1)(n-2)} \label{a1a2}  \\
&e_1=\frac{1}{(n-2)} \quad , \quad   e_2=-\frac{1}{(n-1)(n-2)} \label{b1b2}
\end{align}
allowing us to rewrite the tensors $G_{ijkl}$ and $T_{ijkl}$ as: 
\begin{align}
G_{ijkl}= & \, a C_{ijkl} + B_{ijkl}, \\
T_{ijkl} = & \, e T^{(C)}_{ijkl} + T^{(M)}_{ijkl}.   
\end{align} 
By this simple calculation, we quickly find here the tensors $B_{ijkl}$ and  $T^{(M)}_{ijkl}$,  already defined in (\ref{Bijkl}) and  (\ref{TMijkl}) and, by a change of notation, we can set $eT^{(C)}_{ijkl}=T^{(F)}_{ijkl}$ in accordance with the notation first used  in (\ref{TijklTMTF}), $T_{ijkl}=T^{(F)}_{ijkl} + T^{(M)}_{ijkl}$.    
\\  
To ensure the conservation of the total energy-momentum tensor $\nabla_{i}\, T^{i}{}_{jkl}=0$, we must have the relationship $\nabla_{i}\, G^{i}{}_{jkl}=0$, giving us the value of the parameter $a$:
\begin{equation}
\nabla _{i} G^{i}{}_{jkl}= - \frac{1 + a(n-3)}{(n-2)} \, C_{jkl}=0 \quad \Rightarrow \quad  a=-\frac{1}{(n-3)} \label{nablaGijkl}
\end{equation}
where we identify here the well-known Cotton tensor \cite{citBlau}:  $ C_{jkl}=\nabla_{l} R_{jk} - \nabla_{k} R_{jl} 
 + {1}{/2(n-1)} (g_{jl}\nabla_{k} R - g_{jk}\nabla_{l} R)$. The generalized fourth-order Einstein tensor $G_{ijkl}$ can therefore be written in the form:
\begin{equation}
 G_{ijkl} =  -\frac{1}{(n-3)} C_{ijkl} + B_{ijkl} =  \frac{1}{(n-3)}\, ^{*}R_{ijkl}^{*},  
\end{equation}
thus giving, with  (\ref{GijklTijkl}), the sought-after generalized 4-index Einstein equation (\ref{eqdual4index}). 
       \subsection{Number of independent components}   \label{section2.5}
The number of algebraically independent components of the various tensors used in this paper, are given in Weinberg's book \cite{citWeinberg} and summarized in the table below:
\begin{center}
\renewcommand{\arraystretch}{1.5}
\begin{tabular}{|p{6.9cm}|p{3cm}|p{2cm}|}
\hline 
 Number of independent components & \textit{n-dimensions} & \textit{n=4}   \\ 
\hline 
$R_{ijkl}$   Riemann tensor  & $\frac{n^2(n^2-1)}{12}$ & 20  \\ 
\hline 
$^{*}R_{ijkl}^{*}$ Riemann double dual tensor     & $\frac{n^2(n^2-1)}{12}$ & 20  \\
\hline 
$g_{jl}$  Metric tensor  & $\frac{n(n+1)}{2} $ & 10  \\
\hline 
$R_{jl}$  Ricci tensor  & $\frac{n(n+1)}{2} $ & 10  \\ 
\hline
$B_{ijkl}$  tensor & $\frac{n(n+1)}{2} $ & 10  \\ 
\hline
$C_{ijkl}$ Weyl tensor  & $\frac{n(n+1)(n+2)(n-3)}{12} $ & 10  \\ 
\hline 
\end{tabular}
\end{center}
In dimension \textit{n=4}, the 20 independent components of the Riemann tensor distribute into the 10 components in the Ricci tensor and 10 components in the Weyl tensor. 
The Einstein equation (\ref{eqeinstein}) contains 10 independent equations, while the general equation (\ref{eqdual4index1}) or (\ref{eqdual4index})  contains 20. The decomposition into the two equations (\ref{BijklTMijkl}) and (\ref{CijklTFijkl}), tells us that for \textit{n=4}, the additional 10 degrees of freedom are therefore  contained in the new equation $C_{ijkl}=-\chi T^{(F)}_{ijkl}$. 
The Weyl tensor $C_{ijkl}$ is basically the part of the Riemann tensor that is not included in the tensors $R_{jl}$ and $R$, and thus the information contained in (\ref{CijklTFijkl}) is not contained in (\ref{BijklTMijkl}) or in (\ref{eqeinstein}). \\
The ${n(n+1)}/{2}$ components of the metric are completely defined by the usual 2-index Einstein equation and we will see later that the $n(n+1)(n+2)(n-3)/12$ new degrees of freedom, introduced via the 4-index theory, will appear as additional constraints on $g_{jl}$.
        \subsection{The gravitational field tensor $T^{(F)}_{jl}=0$ }  \label{section2.6}
It is possible to decompose the tensor $T_{jl}$ as is done in the relation (\ref{TijklTMTF}):
\begin{equation}
 T_{jl}= T^{(M)}_{jl} + T^{(F)}_{jl},    
\end{equation}
but, an important consequence of the equation (\ref{CijklTFijkl}) is that the tensor $T^{(F)}_{ijkl}$ must have the same properties as a Weyl tensor namely zero contraction:
\begin{equation}
 g^{ik}\, T^{(F)}_{ijkl}=T^{(F)}_{jl}=0,    \label{gikTFijkl}
\end{equation}
and that consequently: 
\begin{equation}
 T_{jl}\equiv T^{(M)}_{jl}  \quad \text{and} \quad T \equiv T^{(M)}.   \label{TMjl} 
\end{equation}
The Einstein equation (\ref{eqeinstein}) cannot  explicitly  contain a 2-index tensor $T^{(F)}_{jl}$ for the gravitational field (however, a possible energy-momentum tensor of an electromagnetic field, $T^{(F)(e)}_{jl}$, may of course be present, but we won't take it into account in our calculations). 
This result may perhaps explain why, despite decades of effort, the search for a symmetrical 2-index tensor or pseudo-tensor for the gravitational field and introduced directly into Einstein's equation (\ref{eqeinstein}) under the form   $\chi (T^{(M)}_{jl} + T^{(F)}_{jl})$ often noted $\chi (T^{(M)}_{jl} + t_{jl})$, has, in most cases, not given complete satisfaction (the Einstein pseudo-tensor and the Landau-Lifchitz pseudo-tensor $t_{jl}$ are among the best known examples \cite{citLandau,citSzabados,citFavata,citHobson}).\\
The 4-index approach proposed here, with the natural introduction of the traceless tensor $T^{(F)}_{ijkl}$, is necessary to complete the relativistic mathematical formulation of the gravitational field. 
        \subsection{Determining the gravitational field tensor $T^{(F)}_{ijkl}$ }  \label{section2.7}
For the purposes of the calculations that follow, we will restrict ourselves to the simple case of a gravitational field possessing central symmetry.  We will use a general diagonal metric tensor $g_{jl}$, written in a spherical coordinate system $(ct, r, \theta, \phi)$ whose components will be denoted $(0, 1, 2, 3)$ in a space-time of dimension \textit{n=4} : 
\begin{align}
ds^2= g_{jl}\,dx^j dx^l=g_{00} \,c^2 dt^2 + g_{11} \,dr^2 + g_{22} \, (d\theta^2 +  sin^2\theta d\phi^2)  \nonumber
\end{align}
Einstein's equation (\ref{eqeinstein}) with the relation  (\ref{TMjl}), and the general equation (\ref{eqdual4index}) with the relation  (\ref{TijklTMTF}), can be rewritten in mixed notation as (the mixed tensor notation $(^{j}{}_{l})$ or $(^{ij}{}_{kl})$ are more suitable for the calculations):
\begin{align}
 G^{j}{}_{l}=&\chi \, T^{(M)j}{}_{l}, \label{Einsteinmixte} \\
{^{*}}R^{*\,ij}{}_{kl}   =& \chi  T^{(M)ij}{}_{kl} + \chi  T^{(F)ij}{}_{kl} ,  \label{genralizedEinsteinmixte} 
\end{align}  
with the Einstein tensor, $G^{j}{}_{l}=R^{j}{}_{l} -{1}/{2}\,\delta^j_l R$, and with the matter tensor $T^{(M)ij}{}_{kl}$ (\ref{TMijkl}), write for \textit{n=4}, with (\ref{TMjl}):
\begin{equation}
T^{(M)ij}{}_{kl}=  \frac{1}{2}  ( \delta^i_k  T^{(M)j}{}_{l} - \delta^j_k T^{(M)i}{}_{l} + \delta^j_l  T^{(M)i}{}_{k}
-\delta^i_l T^{(M)j}{}_{k})   -  \frac{1}{6}  ( \delta^i_k \delta^j_l -\delta^i_l\delta^j_k)T^{(M)}   \label{TMijkl2} 
\end{equation}
We will consider a simple but important case, such as the stress-energy tensor of a perfect fluid for example, given by a diagonal energy-momentum tensor which takes a particularly simple form in the rest frame \cite{citHobson}:
\label{matrice}
\begin{eqnarray} 
             T^{(M)j}{}_{l}=
					\begin{aligned}
					\begin{pmatrix}
\, \rho c^2      & \quad 0         & \quad   0        & \quad  0 \, \\ 
\, 0         & \quad -p        & \quad   0        & \quad  0  \, \\
\, 0         & \quad  0        & \quad   -p       & \quad  0 \, \\                             
 \, 0        & \quad  0        & \quad   0        & \quad  -p \,    \label{Tjlrhop}
              \end{pmatrix}  
          \end{aligned}                             
\end{eqnarray} 
where $\rho$ is the rest frame matter density and $p$ the isotropic pressure.  \\
In the calculations that follow, we will keep the notation:\\ $T^{(M)j}{}_{l}= diag(T^{(M)0}{}_{0}, T^{(M)1}{}_{1}, T^{(M)1}{}_{1}, T^{(M)1}{}_{1})$ allowing us to calculate:    \\
$T^{(M)01}{}_{01}=T^{(M)02}{}_{02}=T^{(M)03}{}_{03}= T^{(M)0}{}_{0}/3$, \\ 
$T^{(M)12}{}_{12}=T^{(M)13}{}_{13}=T^{(M)23}{}_{23}= -T^{(M)0}{}_{0}/6 + T^{(M)1}{}_{1}/2$. \\
The equations (\ref{Einsteinmixte}) and (\ref{genralizedEinsteinmixte}) both contain the energy-momentum tensor of matter $T^{(M)j}{}_{l}$ and $T^{(M)ij}{}_{kl}$ respectively, so their different components must be compatible with respect to $^{(M)}$. 
The calculation technique, that we will use here to determine as simply as possible the form of the tensor $T^{(F)}_{ijkl}$, will consist of comparing these two equations component by component. \\
For example, let's compare the first component, $G^{0}{}_{0}= \chi T^{(M)0}{}_{0}$, of (\ref{Einsteinmixte}):
\begin{equation}
   G^{0}{}_{0} =  \frac{g''_{22}}{g_{11} g_{22}} - \frac{ (g'_{22})^2}{4g_{11} g^2_{22}}
	-\frac{ g'_{11}  g'_{22}}{2g^2_{11} g_{22}} - \frac{ 1}{ g_{22}} =\chi  T^{(M)0}{}_{0},     \label{G00mixte}  
\end{equation}
with the component, $ {^{*}}R^{*\,01}{}_{01}= \chi T^{(M)01}{}_{01} + \chi T^{(F)01}{}_{01}$   of (\ref{genralizedEinsteinmixte})
\begin{equation}
  {^{*}}R^{*\,01}{}_{01} =  \frac{ (g'_{22})^2}{4g_{11} g^2_{22}} - \frac{ 1}{ g_{22}}
	= \chi  T^{(M)01}{}_{01} + \chi T^{(F)01}{}_{01}.
   \label{R0202mixte}
\end{equation}
The first and second spatial derivatives, (${\partial}/{\partial r}$)  and (${\partial^2}/{\partial r^2} $), are noted respectively ($'$) and ($''$).
The relation, $T^{(M)01}{}_{01} =T^{(M)0}{}_{0}/3$, tells us that the two expressions (\ref{G00mixte}) and  (\ref{R0202mixte}) are compatible with each other if we set, by identification:
\begin{equation}
 \chi T^{(F)01}{}_{01}  = -\frac{g''_{22}}{3g_{11} g_{22}} + \frac{ (g'_{22})^2}{3g_{11} g^2_{22}}
	+\frac{ g'_{11}  g'_{22}}{6g^2_{11} g_{22}} - \frac{ 2}{ 3g_{22}}.  \label{quiT0202mixte}
\end{equation}
Surprising as it may seem, the above expression is simply given by calculating the Weyl tensor component, $-2C^{\,01}{}_{01}$, obtained with the formula (\ref{Weytensor}) for $n=4$ and for fully mixed tensor notation, simply by forcing the time metric component $g_{00}=1$. After a component-by-component comparison, so that all components of the 2- and 4-index equations (\ref{Einsteinmixte}) and (\ref{genralizedEinsteinmixte}) are fully compatible with each other, we find a fairly simple mathematical solution for $T^{(F)ij}{}_{kl}$ : 
\begin{equation}
 T^{(F)ij}{}_{kl}   = - \frac{2}{\chi}\, [\,C^{\,ij}{}_{kl}\,]_{g_{00}=1}      \label{deftijkl2}  
\end{equation}
How can we interpret this result ? \\
It is difficult to give a simple physical meaning to this Weyl tensor $[\,C^{\,ij}{}_{kl}\,]_{g_{00}=1}$ which seems to be calculated in a synchronous frame. The particular relation (\ref{deftijkl2}) has not been posed and seems to be mathematically essential in order to be compatible at least with the usual theory of general relativity.
Let us also recall that the calculations were carried out within the framework of a time-independent gravitational field and, in the case of the Schwarzschild metric in vacuum (see part \ref{section3.4}), the calculation will give a very simple result, $[\,C^{\,ij}{}_{kl}\,]_{g_{00}=1} \propto {1}/{ r^3} $,  showing a well-known result, namely that the Weyl tensor is closely related to the tidal effects of gravity \cite{citEllis2,citEllis,citBertschinger}. \\
As stated in some references, such as in the book \cite{citOyv}: "The Weyl tensor represents the free gravitational field" and in \cite{citEllis2}: "Roger Penrose suggested that it (the free gravitational field) should be constructed entirely via the Weyl tensor, which describes the free gravity on the manifold". We can see that the calculations presented here are in line with what many physicists believe, keeping in mind that strictly speaking, $T^{(F)}_{ijkl}$ is not directly the tensor of the gravitational field itself but rather the tensor of the tidal gravitational field. \\
Future research could certainly, I hope, lead to a general mathematical resolution, regardless of the form of the metric or independently of the form of the energy-momentum tensor.
  \section{ Schwarzschild metric}   \label{section3}
Let us now look at what these last relationships give on well-known metrics. The starting space-time interval required to obtain the Schwarzschild metric is generally written as \cite{citHobson}:
\begin{align}
ds^2= A(r) \,c^2 dt^2 - B(r) \,dr^2 - r^2 \, (d\theta^2 +  sin^2\theta d\phi^2),   
\end{align}
where we have chosen a $(+---)$ signature with the spherical coordinates $(ct, r, \theta, \phi)$ whose components will be noted $(0, 1, 2, 3)$.
  \subsection{ The equation $C^{ij}{}_{kl} =- \chi T^{(F)ij}{}_{kl}$ }  \label{section3.1}
From the definition (\ref{Weytensor}) with $n=4$, we can easily calculate the two main components of the Weyl tensor $C^{ij}{}_{kl}$:
\begin{align}
  &C^{01}{}_{01} =  \, \frac{A''}{6AB} - \frac{A'^2}{12A^2B} - \frac{A'B'}{12AB^2} - \frac{A'}{6rAB} 
 + \, \frac{B'}{6rB^2} + \frac{1}{3r^2B} -\frac{1}{3r^2}  \label{C1212} \\
 &C^{02}{}_{02}=     -\frac{A''}{12AB} + \frac{A'^2}{24A^2B} + \frac{A'B'}{24AB^2} + \frac{A'}{12rAB}  
  -\frac{B'}{12rB^2} - \frac{1}{6r^2B} +\frac{1}{6r^2}  \label{C1313} 
\end{align} 
The other non-zero terms are given by: $C^{01}{}_{01}=C^{23}{}_{23}$ and $ C^{02}{}_{02}=C^{03}{}_{03}=C^{12}{}_{12}=C^{13}{}_{13}$,  and also by symmetries: $C^{ij}{}_{kl}=-C^{ji}{}_{kl}=-C^{ij}{}_{lk}=C^{ji}{}_{lk}$. \\
On the other hand, the tensor $T^{(F)ij}{}_{kl}$ is given by the relation (\ref{deftijkl2}):  
\begin{align}
& T^{(F)01}{}_{01}= \frac{1}{\chi} \left[ -\frac{B'}{3rB^2} -\frac{2}{3r^2B} +\frac{2}{3r^2} \right],  \label{t1212} \\ 
& T^{(F)02}{}_{02}= \frac{1}{\chi} \left[ \frac{B'}{6rB^2} +\frac{1}{3r^2B} -\frac{1}{3r^2} \right].   \label{t1313}
\end{align} 
The other non-zero terms are given by: $T^{(F)01}{}_{01}=T^{(F)23}{}_{23}$ and $ T^{(F)02}{}_{02}=T^{(F)03}{}_{03}=T^{(F)12}{}_{12}=T^{(F)13}{}_{13}$,  and also by symmetries: $T^{(F)ij}{}_{kl}=-T^{(F)ji}{}_{kl}=-T^{(F)ij}{}_{lk}=T^{(F)ji}{}_{lk}$. \\
The equation (\ref{CijklTFijkl}) is now written for a space-time of dimension \textit{n=4}:
\begin{equation}
C^{ij}{}_{kl} + \chi T^{(F)ij}{}_{kl}=0, \label{CijklTFijkl2}
\end{equation}
whose main components are given by:
\begin{align}
   C^{01}{}_{01} + \chi T^{(F)01}{}_{01}  = &  \,\, \frac{A''}{6AB} - \frac{A'^2}{12A^2B} - \frac{A'B'}{12AB^2}     
	 - \frac{A'}{6rAB}-\frac{B'}{6rB^2}  - \frac{1}{3r^2B} +\frac{1}{3r^2}=0,  \\
 C^{02}{}_{02} + \chi T^{(F)02}{}_{02} = & -\frac{A''}{12AB} + \frac{A'^2}{24A^2B} + \frac{A'B'}{24AB^2}   
+ \frac{A'}{12rAB} +\frac{B'}{12rB^2}  + \frac{1}{6r^2B}-\frac{1}{6r^2}=0.   
\end{align} 
Surprising as it may seem, we notice that all the components give the same relationship except for one constant which can ultimately be simplified. We will write this new relationship, obtained here for the first time, in a form that will be useful for subsequent calculations: 
\begin{align}
&\frac{A''}{2AB} - \frac{A'^2}{4A^2B} - \frac{A'B'}{4AB^2}-  \frac{A'}{2rAB} -\frac{B'}{2rB^2} 
- \frac{1}{r^2B} + \frac{1}{r^2} =0.   \label{relationconstB}
\end{align} 
The 4-index equation (\ref{CijklTFijkl2}), gives a new additional constraint relation to which the terms $A(r)$ and $B(r)$ of the metric must obey. The relation (\ref{relationconstB})  is valid for both the Schwarzschild exterior and interior cases. 
  \subsection{ Exterior metric with $C^{ij}{}_{kl} =- \chi T^{(F)ij}{}_{kl}$ }  \label{section3.2}
For the Schwarzschild exterior case, the gravitational field equations in vacuum, gives the well-known starting relation, $B(r)=1/A(r)$ \cite{citLandau,citMisner}, allowing us to simplify the equation (\ref{relationconstB}) as: 
\begin{align}
 \frac{A''}{r}   -  \frac{2A}{r^3} +  \frac{2}{r^3} =0 \label{relationA}   
\end{align}  
which can also be written: 
\begin{align}
 ( \frac{A'}{r}   +  \frac{A}{r^2} -  \frac{1}{r^2} )' =0   
\end{align}
giving:
\begin{align}
\frac{A'}{r}   +  \frac{A}{r^2} -  \frac{1}{r^2} + \lambda =0
\end{align}
or:
\begin{align}
(r A)'  -1 + \lambda r^2=0
\end{align}
having the solution:
\begin{align}
A(r)= 1 - \frac{k}{r} -  \frac{\lambda r^2}{3} \label{Kottlera}
\end{align}
where $k$ and $\lambda$ are integration constants. \\
We immediately notice that this solution has the same form as the Schwarzschild de-Sitter solution (or Kottler metric \cite{citBlau}):
\begin{align}
A(r)= \frac{1}{B(r)} =  1 - \frac{r_s}{r} -  \frac{\Lambda r^2}{3},   \label{Kottler}
\end{align}
which, as everyone knows, was obtained by introducing ad hoc the cosmological constant $\Lambda$ directly into Einstein's equations. \\
The 4-index equation (\ref{CijklTFijkl2}) therefore naturally gives both the usual solution of the gravitational field potential, $(\propto 1/r)$, produced by the presence of matter, but at the same time, spontaneously, a new solution $(\propto r^2)$, without the need to introduce any constant $\lambda$ into the equations or into the Lagrangian. 
The usual interpretation of $\Lambda$ in terms of dark energy \cite{citPeebles} will not necessarily be that of $\lambda$ which appears here as an integration constant.
The additional degrees of freedom introduced by equation (\ref{CijklTFijkl2}), clearly show us that we cannot exclude the possibility that the two physical solutions, $(k/r)$ and $(\lambda r^2/3)$, are somehow closely linked by some as yet unknown property, or physical constraint, concerning the gravitational field. \\
For simplicity of notation in the remainder of the article, we will retain the same notations and names as those used in the Schwarzschild de-Sitter case  by setting  the Schwarzschild radius $k=r_s$ and the cosmological constant $\lambda=\Lambda$, bearing in mind that the usual interpretation of $\Lambda$ will not necessarily be that of $\lambda$.
  \subsection{ Interior metric with $C^{ij}{}_{kl} =- \chi T^{(F)ij}{}_{kl}$ }  \label{section3.3}
The interior Schwarzschild metric is an exact solution for the gravitational field of a non-rotating spherical body with a constant density $\rho$. In this specific case, it is well known and demonstrated in many books, \cite{citMisner,citOyv} that $B(r)$ for the interior case, can be written in a general form, $B(r) = (\, 1- b \, r^2 \, )^{-1}$ with a parameter $b$ to be determined here. The relation (\ref{relationconstB}) can then be written as: 
\begin{align}
\frac{A''}{A'} - \frac{A'}{2A} -  \frac{1}{r} - \frac{b\,r}{1- b\,r^2} =0  \label{relationconstBinterior}
\end{align} 
which we integrate to obtain:
\begin{align}
\frac{A'}{\sqrt{A}} =  \frac{K_1 r}{ \sqrt{1- b\,r^2}}  
\end{align} 
having the solution:
\begin{align}
A(r) = \left[\, \frac{K_2}{2} - \frac{K_1 }{2b} \sqrt{1- b\,r^2} \, \right]^2  \label{Ageneral}
\end{align} 
where $K_1$ and $K_2$ are integration constants. \\
For the value of the r-coordinate at the surface of the body $r=R$, we will use the usual boundary conditions of continuity for $A(r)$, $B(r)$ as well as for the pressure $p(R)$, giving us, $K_1=(b-\Lambda)$ and  $ K_2=(3- \Lambda/b)  \sqrt{1- b\,R^2}$ with $b=(r_s /R^3 + \Lambda/3)  =(\chi \rho c^2/3 + \Lambda/3)$, allowing us to finally write the solution for the interior case:
\begin{align}
 A(r) =&  \bigg[\, \frac{3 \chi\rho c^2}{2( \chi\rho c^2 + \Lambda)}  \sqrt{1- (\frac{\chi\rho c^2}{3} + \frac{\Lambda}{3}) \, R^2}  - \frac{\chi\rho c^2-2\Lambda }{2(\chi\rho c^2 + \Lambda)} \sqrt{1- (\frac{\chi\rho c^2}{3} + \frac{\Lambda}{3}) \, r^2}  \, \bigg]^2 \label{Alambda} \\
 B(r) =& \biggl[\,  1- (\frac{\chi\rho c^2}{3} + \frac{\Lambda}{3}) \, r^2   \, \biggr]^{-1}   \label{Blambda} 
\end{align} 
For $\Lambda=0$, we find the known  solution for the Schwarzschild interior case \cite{citOyv,citHobson}: 
\begin{align}
& A(r) =  \biggl[\, \frac{3}{2}  \sqrt{1- \frac{\chi\rho c^2}{3}\, R^2} - \frac{1}{2} \sqrt{1- \frac{\chi\rho c^2}{3} \, r^2} \,\, \biggr]^2  \\
& B(r) = \biggl[\, 1- \frac{\chi\rho c^2}{3} \, r^2 \, \biggr]^{-1}  \label{Bsanslambda}
\end{align} 
We have therefore shown  that the fundamental equation (\ref{CijklTFijkl2}), gives both Schwarzschild exterior and interior solutions quite easily, naturally including the constant $\Lambda$ (or $\lambda$). 
  \subsection{Components of the tensor $T^{(F)ij}{}_{kl}$}   \label{section3.4}
We can now introduce the expression, $B(r)=(1-{r_s}/r - \Lambda r^2/3)^{-1}$, into relations (\ref{t1212}) and (\ref{t1313}) which allows us to calculate the two main components of the tensor $T^{(F)ij}{}_{kl}$ for the Schwarzschild metric exterior case:
\begin{align}
& T^{(F)01}{}_{01}=\frac{1}{\chi} (\frac{r_s}{ r^3} )  \label{t1212bis}  \\ 
& T^{(F)02}{}_{02}=  \frac{1}{\chi} (-\frac{r_s}{2 r^3} )  \label{t1313bis}
\end{align} 
The $(1/r^3)$ dependence clearly demonstrates that the tensor $T^{(F)ij}{}_{kl}$ represents a tidal gravitational field \cite{citMisner,citBlau,citHobson}. The constant $\Lambda$ (or $\lambda$) has a null contribution in $T^{(F)ij}{}_{kl}$ and is therefore not associated with any tidal gravitational field but this does not, of course, exclude the possibility of its being at the origin of a particular property of the gravitational field.
  \subsection{Components of the tensor $T^{(M)ij}{}_{kl}$}  \label{section3.5}
The two equations (\ref{BijklTMijkl}) and (\ref{CijklTFijkl}) are coupled via the metric and its derivatives and must therefore be compatible with respect to the the physical solutions, $A(r)$ and $B(r)$, obtained previously. 
A detailed mathematical study of the equation (\ref{BijklTMijkl}), $B_{ijkl}=\chi T^{(M)}_{ijkl}$, indicates to us that the total 4-index energy-momentum tensor $T^{(M)ij}{}_{kl}$, must then be formed by two parts as discussed in section \ref{section3.2}:
\begin{equation}
  T^{(M)ij}{}_{kl}= T^{(M_G)ij}{}_{kl}+T^{(M_\Lambda)ij}{}_{kl}.  \label{TMLambda}
\end{equation}
The first, $T^{(M_G)ij}{}_{kl}$, concerns the well-known usual part related to the stress-energy tensor of a perfect fluid which can be easily calculated using (\ref{TMijkl2}) and (\ref{Tjlrhop}). The second part, $T^{(M_\Lambda)ij}{}_{kl}$, can be determined by also studying the gravitational field equation (\ref{BijklTMijkl}), using the integration constant $\lambda=\Lambda$  such that:  
\begin{equation}
  T^{(M_\Lambda)ij}{}_{kl}=\frac{1}{3 \chi}   \, (\, \delta^i_k \delta^j_l -\delta^i_l \delta^j_k \,) \Lambda. \label{Tlambda}
\end{equation}
The contraction of (\ref{TMLambda}) gives:
\begin{equation}
 T^{(M)j}{}_{l}= T^{(M_G)j}{}_{l} + T^{(M_\Lambda)j}{}_{l},
\end{equation}
with $T^{(M_\Lambda)j}{}_{l}={1}/{ \chi} \, \delta^j_l \Lambda$ which is therefore now automatically included into Einstein's equation. 
As already indicated in parts \ref{section3.2} and \ref{section3.3}, we find here  the two closely related components of the gravitational field. The presence of $\Lambda$  is not optional and will always be part of the tensor $T^{(M)ij}{}_{kl}$ or $T^{(M)j}{}_{l}$. 
  \subsection{ Einstein's equation components for the Schwarzschild metric}  \label{section3.6}
The components of Einstein's equation, $G^{j}{}_{l}=R^{j}{}_{l} -1/2 \, \delta^j_l R=\chi  T^{j}{}_{l}$, are given by:
\begin{align}
 G^{0}{}_{0}= &\, \frac{B'}{rB^2} - \frac{1}{r^2B} +\frac{1}{r^2} = \chi  T^{0}{}_{0},   \label{EinsteinG11} \\
 G^{1}{}_{1}= &-\frac{A'}{rAB} - \frac{1}{r^2B} +\frac{1}{r^2} = \chi  T^{1}{}_{1},   \label{EinsteinG22}  \\
 G^{2}{}_{2}= & -\frac{A''}{2AB} + \frac{A'^2}{4A^2B} + \frac{A'B'}{4AB^2}  - \frac{A'}{2rAB} + \frac{B'}{2rB^2} 
   =   \chi  T^{2}{}_{2}, \label{EinsteinG33} \\
 G^{3}{}_{3}= & -\frac{A''}{2AB} + \frac{A'^2}{4A^2B} + \frac{A'B'}{4AB^2}- \frac{A'}{2rAB}  + \frac{B'}{2rB^2} 
   =  \chi  T^{3}{}_{3}. \label{EinsteinG44}
\end{align} 
Let us see what the relation (\ref{relationconstB}) gives when we introduce it into Einstein's equation (\ref{EinsteinG33}) (or (\ref{EinsteinG44})): 
\begin{align}
 G^{2}{}_{2}=G^{3}{}_{3}= -\frac{A'}{rAB} - \frac{1}{r^2B} +\frac{1}{r^2}, \label{EinsteinG55}
\end{align} 
which becomes identical to Einstein's equation $G^{1}{}_{1}$, giving: 
\begin{align}
G^{1}{}_{1}=G^{2}{}_{2}=G^{3}{}_{3}, \label{T11T22T33}
\end{align} 
and therefore, $T^{1}{}_{1}=T^{2}{}_{2}=T^{3}{}_{3}$.  Here, we find the case already studied of the stress-energy tensor of a perfect fluid (\ref{Tjlrhop}) , given by a diagonal energy-momentum tensor, $ T^{(M)j}{}_{l}=diag(T^{(M)0}{}_{0}, T^{(M)1}{}_{1}, T^{(M)1}{}_{1}, T^{(M)1}{}_{1})$.
This confirms that the equation (\ref{CijklTFijkl2}) is fully compatible with general relativity, even simplifying some of Einstein's equations and eliminating spatial second derivatives in the process. 
  \subsection{ Generalized 4-index Einstein's equation components for the Schwarzschild metric}  \label{section3.7}
For the exterior Schwarzschild case, the tensor of the perfect fluid in vacuum cancels out, $T^{(M_G)ij}{}_{kl}=0$, and then, $T^{(M)ij}{}_{kl}= T^{(M_\Lambda)ij}{}_{kl}$ is given by (\ref{Tlambda}). The gravitational field tensor, $T^{(F)ij}{}_{kl}$ is given by the components (\ref{t1212bis}) and (\ref{t1313bis}) and the independent components of the generalized 4-index equation can therefore be written as follows: 
\begin{align}
^{*}R^{*ij}{}_{kl}  &= \chi T^{(M_\Lambda)ij}{}_{kl} + \chi T^{(F)ij}{}_{kl} \quad \overset{\textit{(exterior Schwarzschild case)}} \nonumber \\
   - \frac{1}{r^2B} \,+\, \frac{1}{r^2} & = \frac{\Lambda}{3} \,+\, \frac{r_s}{r^3} \quad\quad \overset{(^{01}{}_{01})}{}  \label{RDD1212}  \\
    \frac{B'}{2rB^2} & = \frac{\Lambda}{3} \,-\, \frac{r_s}{2r^3} \quad\quad \overset{(^{02}{}_{02})}{}     \\
 -\frac{A'}{2rAB} & = \frac{\Lambda}{3} \,-\, \frac{r_s}{2r^3}    \quad\quad  \overset{(^{12}{}_{12})}{}    
\end{align}  
Solving these various equations easily yields directly the expected solutions for the Schwarzschild de-Sitter case   (\ref{Kottler}). The free (tidal) gravitational field in vacuum now appears clearly in the 4-index theory through the terms in $1/r^3$. \\
The solution for the interior Schwarzschild case can also be calculated with the equation (\ref{genralizedEinsteinmixte}) by introducing the matter density $\rho$ and the isotropic pressure $p(r)$ inside $T^{(M_G)ij}{}_{kl}$. 
  \section{Friedmann-Lemaître-Robertson-Walker metric}  \label{section4}
We start from the well-known general relation for the Friedmann-Lemaître-Robertson-Walker (FLRW) metric \cite{citHobson}:
\begin{align}
ds^2= c^2 dt^2 - a^2(t) \big[ \, B(r)\,dr^2 + r^2 d\theta^2 +  r^2 \,sin^2\theta d\phi^2\,\big]  \label{FRLWmetricB}  
\end{align}
with $a(t)$ being the scale factor of the space-time and with $B(r)= 1/(1 - q\,r^2)$, where $q$ may be taken to belong to the set, $\left\{-1, 0, +1\right\}$ for negative, zero, and positive curvatures respectively. 
For the calculations that follow, we will leave the metric (\ref{FRLWmetricB}) as it is for the moment with the function $B(r)$. \\ 
Since we have the term $g_{00}=1$, we have the equality for the Weyl tensors, ${C}^{ij}{}_{kl}=[\,C^{\,ij}{}_{kl}\,]_{g_{00}=1}$.
The equation  (\ref{CijklTFijkl2}) is therefore simply equivalent to the nullity of the Weyl tensor:   
\begin{align}
C^{ij}{}_{kl}= 0   \label{Cijkl0} 
\end{align}
The latter result is sometimes written as a consequence of the FLRW metric \cite{citBlau}, but here, using a rigorous mathematical treatment based on the principle of least action, we show that $C^{ij}{}_{kl}= 0$ is a true equation that the FLRW metric must satisfy.
A quick calculation shows that all the components of the Weyl tensor give the following equivalent relationship:
\begin{align}
C^{ij}{}_{kl}= 0 \quad\,\, \Rightarrow \quad\,\, \frac{B'}{2rB^2} + \frac{1}{r^2B} - \frac{1}{r^2} =0  
\end{align} 
which can be simply rewritten as: $(1/r^2B)'+ 2/r^3=0$, giving the solution, $B(r)= 1/(1 - b\,r^2)$ where $b$ is an integration constant. \\
The 4-index equation (\ref{CijklTFijkl2}) therefore gives a solution in agreement with the FLRW metric. 
\section{Conclusion}
In this paper, using the principle of least action as well as a direct determination, we have shown that there exists a 4-index version of the Einstein equation in a \textit{n}-dimensional space that includes the Riemann and Weyl tensors explicitly and linearly, as well as a tidal gravitational field tensor. \\
The article highlights some interesting results, the first of which is given by this general gravitational field equation which can be written with the double dual Riemann tensor and whose total energy is conserved:
\begin{eqnarray} 
					\left\lbrace 
					\begin{aligned}
					& ^{*}R_{ijkl}^{*} = \chi (n-3) \, (\, T^{(M)}_{ijkl} + T^{(F)}_{ijkl}\,)  \\
          &  \nabla_{i} (\,  T^{(M)}{}^{i}{}_{jkl}+T^{(F)}{}^{i}{}_{jkl}\,)=0 
          \end{aligned} \right.  \nonumber                        
\end{eqnarray}
where $T^{(M)}_{ijkl}$ is the energy-momentum tensor of matter, and $T^{(F)}_{ijkl}$ the tidal gravitational field tensor. \\ 
We have shown that this equation can be split into two coupled 4-index equations, and, for a space-time of dimension \textit{n=4}, the first is equivalent to the 2-index Einstein equation, and the second concerns the Weyl tensor $C_{ijkl}$ physically linked to the tidal gravitational field:
\begin{eqnarray} 
					\left\lbrace 
					\begin{aligned}
					& \, R_{jl}-\frac{1}{2}\,g_{jl}R=\chi  T^{(M)}_{jl} \quad\quad\quad \\
           &  \, C_{ijkl} =- \chi T^{(F)}_{ijkl} \quad\quad\quad 
          \end{aligned} \right.  \nonumber                          
\end{eqnarray}
In terms of independent components, for $n=4$, we see that the equation with $^{*}R_{ijkl}^{*}$ contains 20, which are divided into 10 components for the Einstein equation, and 10 components for the equation written with $C_{ijkl}$. The 10 components of the metric $g_{jl}$ are  completely defined by the Einstein equation and we have shown that the 10 new degrees of freedom, introduced via the Weyl tensor, have been interpreted as an additional constraint on the metric.
Because of its zero contraction, the 4-index equation with $C_{ijkl}$, and more specifically the traceless tensor $T^{(F)}_{ijkl}$, highlights the fact that Einstein's equation cannot explicitly contain a 2-index tensor $T^{(F)}_{jl}$ for the gravitational field, which also justifies the introduction of a fourth-order equation. 
The second equation therefore contains information not found in Einstein's equation and can be considered as a equation completing the theory of general relativity. \\
The four-index equation was solved in the case of the usual Schwarzschild metric, and yielded a solution containing a first classical term for the gravitational field potential but also, at the same time, a second term containing an integration constant $\lambda$ which appears as a yet to be defined particular property of the gravitational field.
This solution has the same form as the Schwarzschild de-Sitter solution which, as everyone knows, was obtained by introducing ad hoc the cosmological constant $\Lambda$ directly into Einstein's equations. The usual interpretation of $\Lambda$ in terms of dark energy added directly into the equations is not necessarily that of $\lambda$. \\
The next stage of our research will involve a closer examination of the role and physical interpretation of the cosmological constant that appears in the various equations.
We will also apply the 4-index equation to other well-known metrics, such as black holes and gravitational waves. 
\section*{Acknowledgments}
The author would like to thank the Reviewers for their thoughtful comments and efforts towards improving the manuscript.
%
%
%
%
%
%

%
%
\end{document}